\documentclass[sigconf,nonacm]{acmart}
\AtBeginDocument{%
  }

\usepackage{algorithmic}
\usepackage{graphicx}
\usepackage{textcomp}
\usepackage{xcolor}
\usepackage{graphicx}
\usepackage{subcaption}

\begin{document}

\title{CryoZip: An Efficient Cryogenic Compressor for Quantum Error Correction Syndromes}





\author{Guanchen Tao}
\affiliation{%
  \institution{University of Michigan}
  \city{Ann Arbor}
  \state{MI}
  \country{USA}}

\author{Alexander Knapen}
\affiliation{%
  \institution{University of Michigan}
  \city{Ann Arbor}
  \state{MI}
  \country{USA}}

\author{Jacob Mack}
\affiliation{%
  \institution{University of Michigan}
  \city{Ann Arbor}
  \state{MI}
  \country{USA}}

\author{Gokul Subramanian Ravi}
\affiliation{%
  \institution{University of Michigan}
  \city{Ann Arbor}
  \state{MI}
  \country{USA}}

\author{Qirui Zhang}
\affiliation{%
  \institution{University of Michigan}
  \city{Ann Arbor}
  \state{MI}
  \country{USA}}

\author{Mehdi Saligane}
\affiliation{%
  \institution{Brown University}
  \city{Providence}
  \state{RI}
  \country{USA}}

\author{Dennis Sylvester}
\affiliation{%
  \institution{University of Michigan}
  \city{Ann Arbor}
  \state{MI}
  \country{USA}}


\begin{abstract}
Scaling fault-tolerant quantum computing is increasingly constrained by the limited bandwidth and power budget across the 4\,K-room temperature (RT) interface. We present CryoZip\footnote{Source code available at: https://github.com/ReaLLMASIC/CryoZip}, a cross-stack cryogenic compression framework that cooperates with a lightweight cryogenic quantum error correction (QEC) predecoder to reduce 4\,K-to-RT syndrome transmission under realistic, circuit-level noise. CryoZip targets sparse syndrome vectors with a sliding-window compression architecture sized under strict decoding latency constraints to maximize energy efficiency. We implement and evaluate the design in 22\,nm FDSOI characterized at 4\,K, using vector-based power, performance, and area analysis to obtain realistic hardware data. CryoZip achieves up to 48$\times$ compression---1.8$\times$ higher than state-of-the-art compressors---across various QEC codes while delivering 4-26$\times$ energy savings. When paired with a QEC predecoder, it yields over 14,238$\times$ bandwidth reduction, 
while energy savings rise to 42$\times$ when accounting for realistic QEC interface overheads.


\end{abstract}

\maketitle

\section{Introduction}
Quantum computing promises breakthroughs in chemistry, biology, and large-scale optimization by addressing problems intractable for classical machines. Realizing fault-tolerant quantum computing (FTQC) requires $10^{5}-10^6$ physical qubits and logical error rates near $10^{-15}$ \cite{beverland2022assessing}, necessitating quantum error correction (QEC) to protect logical qubits via redundancy and active error detection.

The surface code \cite{fowler2012surface} is a leading QEC candidate.
However, its poor encoding efficiency makes scaling to a hundred or more logical qubits costly, motivating the use of other higher-efficiency quantum low-density parity-check (qLDPC) codes such as the color code \cite{bombin2006topological} and bivariate bicycle (BB) codes \cite{bravyi2024high}.

Practical deployment of the QEC code remains challenging. Many leading qubit platforms operate at cryogenic temperatures inside dilution refrigerators \cite{iwai2023cryogenic,van2019electronic}, while QEC decoders typically reside at room temperature (RT). This separation demands high-bandwidth 4\,K–RT links that are constrained by limited microwave I/O, added heat load that degrades coherence, and feedback latency that complicates synchronization.

\begin{figure}
    \centering
    \includegraphics[width=\columnwidth]{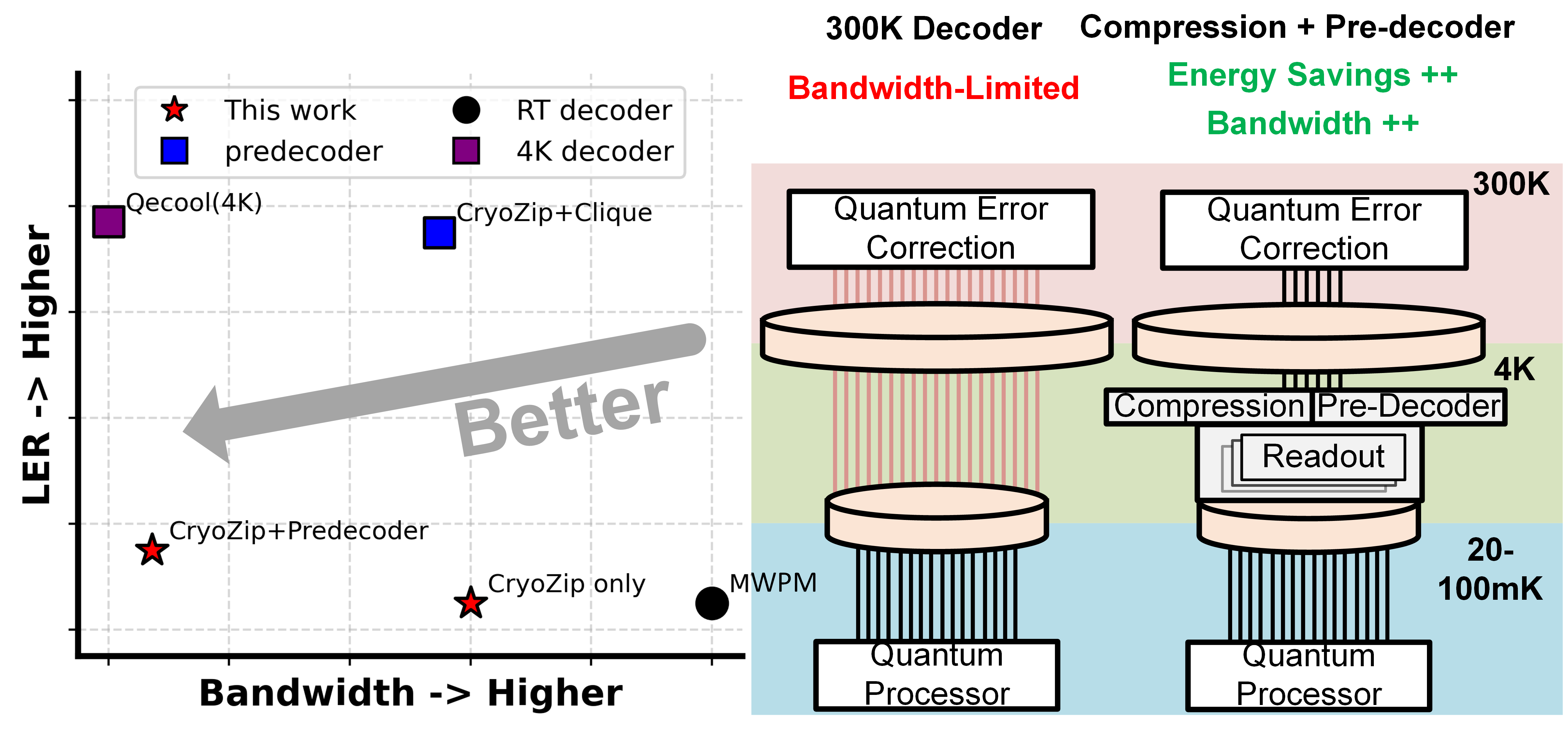}
    \vspace{-0.5cm}
    \caption{
        Landscape of classical hardware solutions for QEC. CryoZip in parallel with predecoder promises bandwidth and energy savings---favorable for bandwidth limited 4\,K quantum computing.
    }
    \label{fig:intro}
    \vspace{-0.5cm}
\end{figure}


To mitigate the control and readout bandwidth bottleneck, both academia and industry have pursued co-locating electronics near the qubits within the cryogenic domain, typically at 4\,K \cite{bardin201929, bardin2019design, chakraborty2022cryo, underwood2024using, van2020scalable, park2021fully, frank2023low, yoo202334}. Thereafter, the bandwidth required for QEC syndrome transmission will increasingly dominate and remain a critical challenge if left unaddressed.

Similar to control and readout, cryogenic decoding has recently been explored, but the $\sim$1.5\,W cooling budget at 4\,K \cite{krinner2019engineering} makes a full-fledged decoder infeasible. Prior proposals for fully cryogenic decoders \cite{holmes2020nisq+,ueno2021qecool,ueno2022qulatis} lack sufficient accuracy and do not scale well to high-distance logical qubits. In contrast, lightweight cryogenic predecoders \cite{ravi2023better, knapen2026pinball} are more practical. These apply local heuristics that exploit sparsity of syndromes, resolving simple errors at low power and flagging complex events for RT decoding. This reduces 4\,K–RT bandwidth while preserving fidelity; however, worst-case events may still require transmitting up to $d$ rounds of syndromes, causing instantaneous power spikes and potential 4\,K–RT backlogs.

To overcome this syndrome challenge, cryogenic syndrome compression co-located with predecoders---or operating independently when confidence is low---can reduce data volume and transmission energy, enabling low-latency, bandwidth-efficient links (Fig.~\ref{fig:intro}). Prior work \cite{das2022afs} explores cryogenic compression in limited fashion - it is evaluated under simplified noise, shows degraded compression under realistic circuit-level noise, and lacks 4\,K hardware validation. Thus, its practicality within dilution refrigerators is uncertain. This motivates a more diligent exploration of cryogenic syndrome compression, co-designed with predecoders under strict 4\,K power limits, with optimization opportunities spanning algorithm, architecture, and technology.

On the technology side, cryogenic CMOS (cryo-CMOS) and single-flux-quantum (SFQ) logic are leading options for low-power cryo-electronics. SFQ offers high speed but faces scaling limits (large area, complex pulse-based design \cite{choi2024supercore}), immature toolchains, and difficult CMOS integration \cite{holmes2020nisq+}. Cryo-CMOS offers greater potential, with high density and mature EDA, and has built significant academic \cite{kang202334,schmidt202513,van2020scalable} and industrial \cite{chakraborty2022cryo,underwood2024using,frank2023low,bardin201929,bardin2019design,yoo202334} interest.

Motivated by these challenges and opportunities, we present \emph{CryoZip}, a cross-layer cryogenic syndrome compression framework spanning algorithm to hardware in 22\,nm FDSOI. CryoZip cooperates with cryogenic predecoders to handle realistic circuit-level noise across diverse QEC codes, while staying within 4\,K power budgets.
The key contributions of this work are:
\begin{enumerate}
    \item \textbf{CryoZip compression algorithm}: We develop a lightweight compression algorithm tailored for sparse syndrome vectors, achieving up to 48.3$\times$ compression. CryoZip delivers up to 1.8$\times$ higher compression compared to state-of-the-art syndrome compression algorithms, enabling bandwidth savings of over 14,238$\times$ when integrated with an existing QEC predecoder \cite{knapen2026pinball} (and 48$\times$ without it). This advantage consistently holds across syndromes generated from three representative QEC codes: surface, BB, and color codes.
    
    \item \textbf{Energy-efficient lightweight hardware design}: By analyzing the stringent latency constraints of QEC decoding, we propose a sliding-window compression architecture and identify the optimal window size that maximizes energy efficiency. With this technique we can achieve up to \textbf{12.2$\times$} area savings and \textbf{8.7$\times$} power savings compared to non-window baseline design. 
    
    \item \textbf{Cryo-CMOS characterization and evaluation}: Using a 22\,nm FDSOI technology re-characterized at 4\,K, we perform vector-based power, performance, and area evaluations of the CryoZip design to obtain realistic hardware data under cryogenic operating conditions. Factoring in cryo-to-RT transmission energy, CryoZip alone achieves 4-26$\times$ energy savings. When combined with the predecoder and accounting for additional overheads from a practical QEC interface \cite{riverlane2024qeci}, the total energy savings increase to 42$\times$.
\end{enumerate}
\section{Background and Motivation}

\subsection{Quantum Error Correction}
QEC uses many error-prone physical qubits (high \textit{physical error rate, $p$}) to redundantly encode one or more robust logical qubit states (low \textit{logical error rate}). Although analogous to classical error correction, QEC is more challenging because qubits suffer both bit ($X$) and phase ($Z$) flips, and directly measuring them collapses their state.

To address this, many QEC codes have been proposed, including quantum low-density parity check (qLDPC) codes. Promising members of this family include surface codes \cite{dennis2002topological, fowler2012surface}, color codes \cite{bombin2006topological}, and bivariate bicycle (BB) codes \cite{bravyi2024high}. In these codes, physical qubits are divided into \textit{data qubits}, which store the logical state, and \textit{ancilla qubits}, which are entangled with data qubits and measured to infer errors indirectly. Ancilla qubits are further divided into $X$- and $Z$-types, which detect $Z$ and $X$ errors on data qubits, respectively.

\begin{figure}
    \centering
    \includegraphics[width=0.8\linewidth]{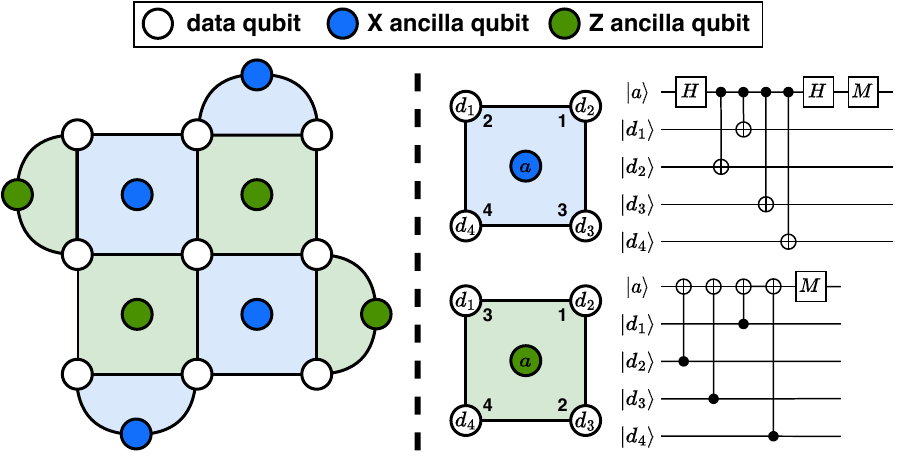}
    \caption{Lattice (left) and syndrome measurement circuit (right) constructions for a distance-3 surface code. Numerical annotations indicate the order of CNOT gates.}
    \label{fig:surface-code}
    \vspace{-0.5cm}
\end{figure}

Fig. \ref{fig:surface-code} illustrates the surface code, the most widely studied QEC code. It uses $2d^2-1$ physical qubits arranged on a 2D lattice, where $d$ is the \textit{code distance}, the minimum number of errors needed to transform one logical qubit state into another. To extract error information from the data qubits (white circles), they are entangled with neighboring ancilla qubits (blue and green circles) using the circuits in Fig. \ref{fig:surface-code}. After each circuit, the ancilla qubits are measured to produce a binary vector called a \textit{syndrome}, which a decoder uses to decide which data qubits to correct. Since syndromes capture only error parity (0 for even and 1 for odd), multiple error patterns can produce the same syndrome.

The surface code is attractive because it requires only nearest-neighbor qubit interactions, but its encoding efficiency scales poorly as $O(d^2)$ physical qubits per logical qubit. Color codes also scale as $O(d^2)$, though with a smaller constant factor, while BB codes scale as $O(d)$ and can encode multiple logical qubits per patch. However, both alternatives require more complex two-qubit interactions.

QEC performance also depends strongly on the assumed error model. In hardware, \textit{all} operations in the circuits of Fig. \ref{fig:surface-code}, as well as qubit idle periods, are error-prone. Realistic evaluation therefore requires a \textit{circuit-level noise model} that captures all of these error sources. Because different hardware platforms exhibit different dominant errors, we use the SI1000 noise model \cite{gidney2021fault}, a circuit-level model motivated by superconducting hardware.

\subsection{QEC Transmission Overhead}
\label{sec:transmission_overhead}
In real QEC systems, decoding must be performed in real time for two reasons. First, if syndromes are produced faster than they are decoded, the backlog grows exponentially \cite{terhal2015quantum}. This is difficult because syndrome measurement latency can be as low as 1 $\mu s$ for superconducting qubits \cite{skoric2023parallel}. Second, some quantum gates, such as $T$ gates, are not fault-tolerant and require corrective gates based on the history of decoded syndromes \cite{fowler2012surface}. The time needed to determine these corrections, called the \textit{response} or \textit{reaction time} \cite{skoric2023parallel, viszlai2025swiper}, also affects application runtime.

Low decoding latency is especially difficult in cryogenic quantum computing systems because transmitting syndrome data from 4 K to RT consumes significant power. For superconducting qubits, the power per syndrome bit is estimated at 0.92 $\mu W$, based on a 1 $\mu s$ syndrome measurement latency and a state-of-the-art cryogenic transmitter energy of 0.92 pJ/b \cite{cusmai20247}. \textit{Even with the full 1.5 W power budget at 4 K} \cite{krinner2019engineering}, only 1000 $d=7$ surface code logical qubits can be supported, far below fault-tolerant computing needs. Passive heat load from cross-temperature cabling further worsens this bottleneck \cite{krinner2019engineering, wang2025wireless}. These limits motivate reducing 4 K-to-RT syndrome bandwidth to enable scalable quantum computing systems.

\begin{figure}
    \centering
    \includegraphics[width=0.8\columnwidth]{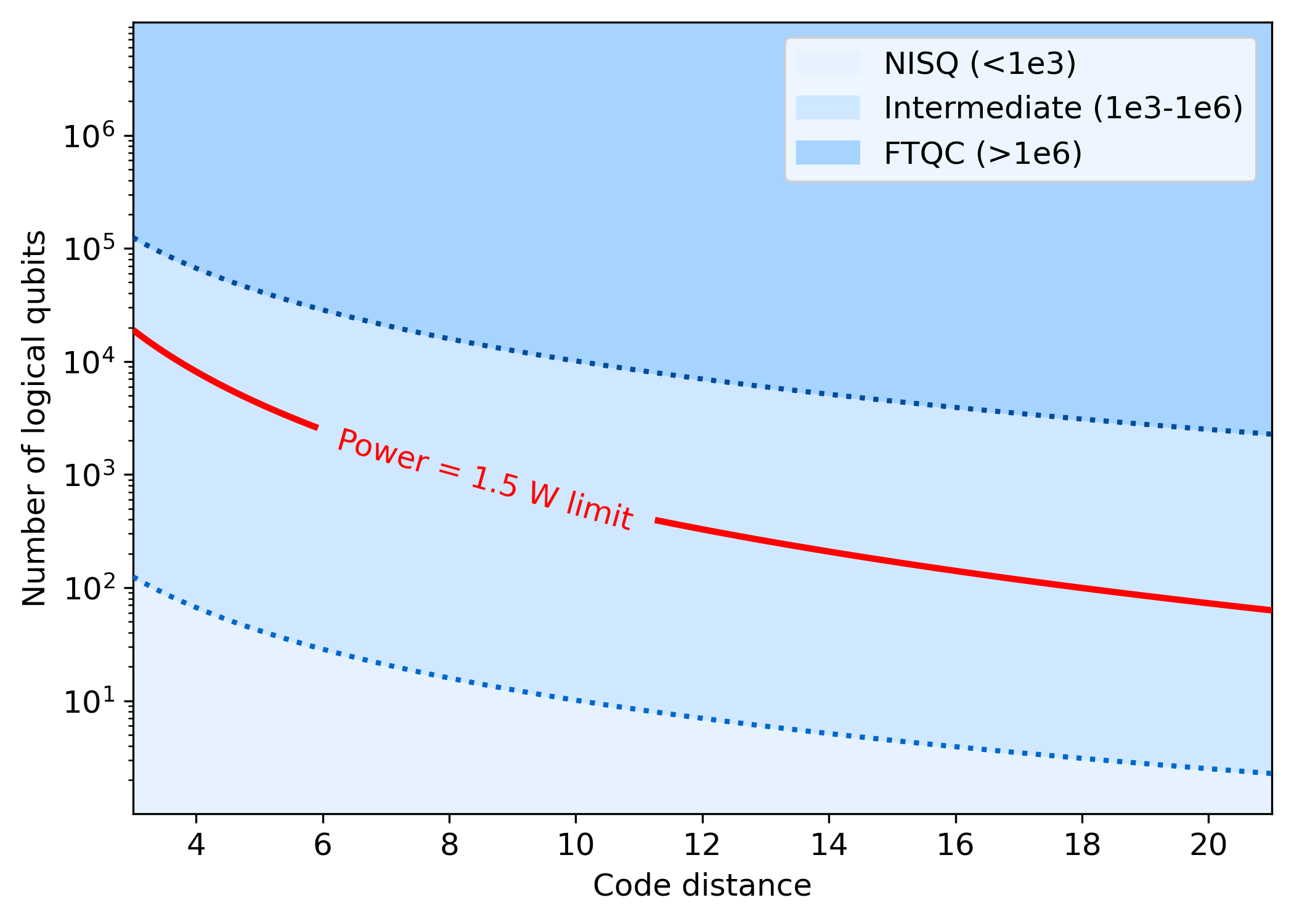}
    \vspace{-1em}
    \caption{Pareto front graph showing the syndrome transmission power needed to support number of logical qubits at various code distances. Red line: transmission power exceeds the 4 K power budget. Blue lines: QEC term according to number of physical qubits.}
    \label{fig:bandwidth}
    \vspace{-0.5cm}
\end{figure}

\section{CryoZip Compression Algorithm}
We begin by introducing CryoZip, a lightweight compression algorithm designed to encode the syndrome vector across $d$ rounds. Existing predecoders reduce average bandwidth by locally correcting trivial, common-case errors. For more complex error patterns beyond their capabilities, complete syndrome vectors must be transmitted to a second-level decoder at room temperature before the cycle ends. While such cases are rare at low error rates and small code distances, they occur more frequently in practical regimes and lead to significant bandwidth spikes. Nevertheless, as syndrome vectors remain sparse even in these scenarios, compression offers a promising solution to reduce bandwidth overhead.

\subsection{Syndrome Compression Pipeline}

\begin{figure}
    \centering
    \includegraphics[width=0.85\columnwidth]{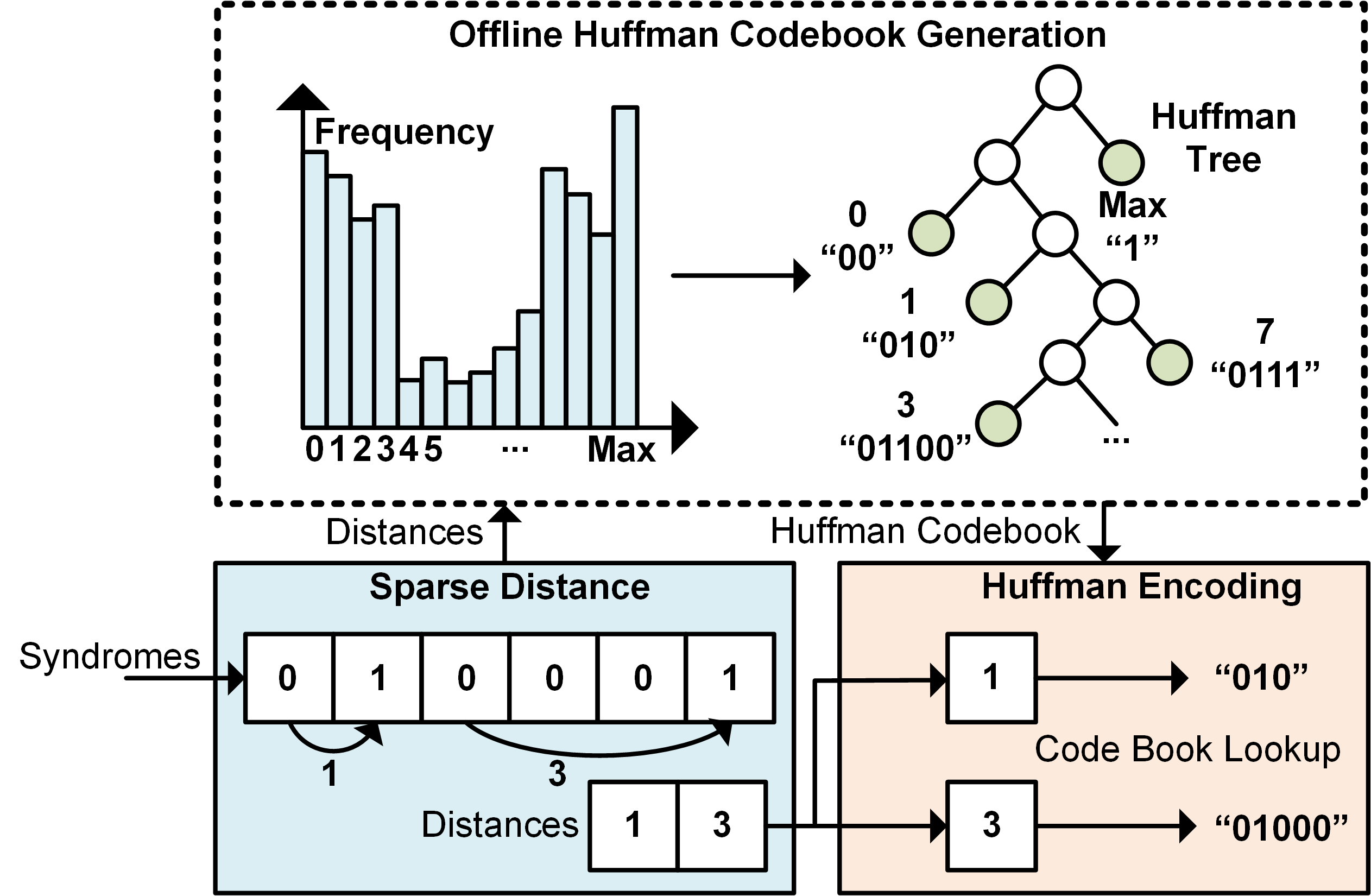}
    \caption{CryoZip algorithm consists of \textit{SD} and \textit{HEnc}. Huffman codebook used for \textit{HEnc} is generated offline.}
    \label{fig:algotop}
    \vspace{-0.25cm}
\end{figure}

CryoZip integrates both our proposed domain-specific compression scheme and classical entropy coding. As illustrated in Fig. \ref{fig:algotop}, the input syndromes first undergo \textit{Sparse Distance (SD)} processing for syndrome reduction, followed by \textit{Huffman Encoding (HEnc)} for bitstream generation. The Huffman codebook is constructed offline by analyzing the statistical frequency of distance values produced by the SD stage, from which a Huffman tree is generated to assign optimal binary codes to each distance symbol. This codebook remains static during the compression.

\begin{figure}
    \centering
    \includegraphics[width=\columnwidth]{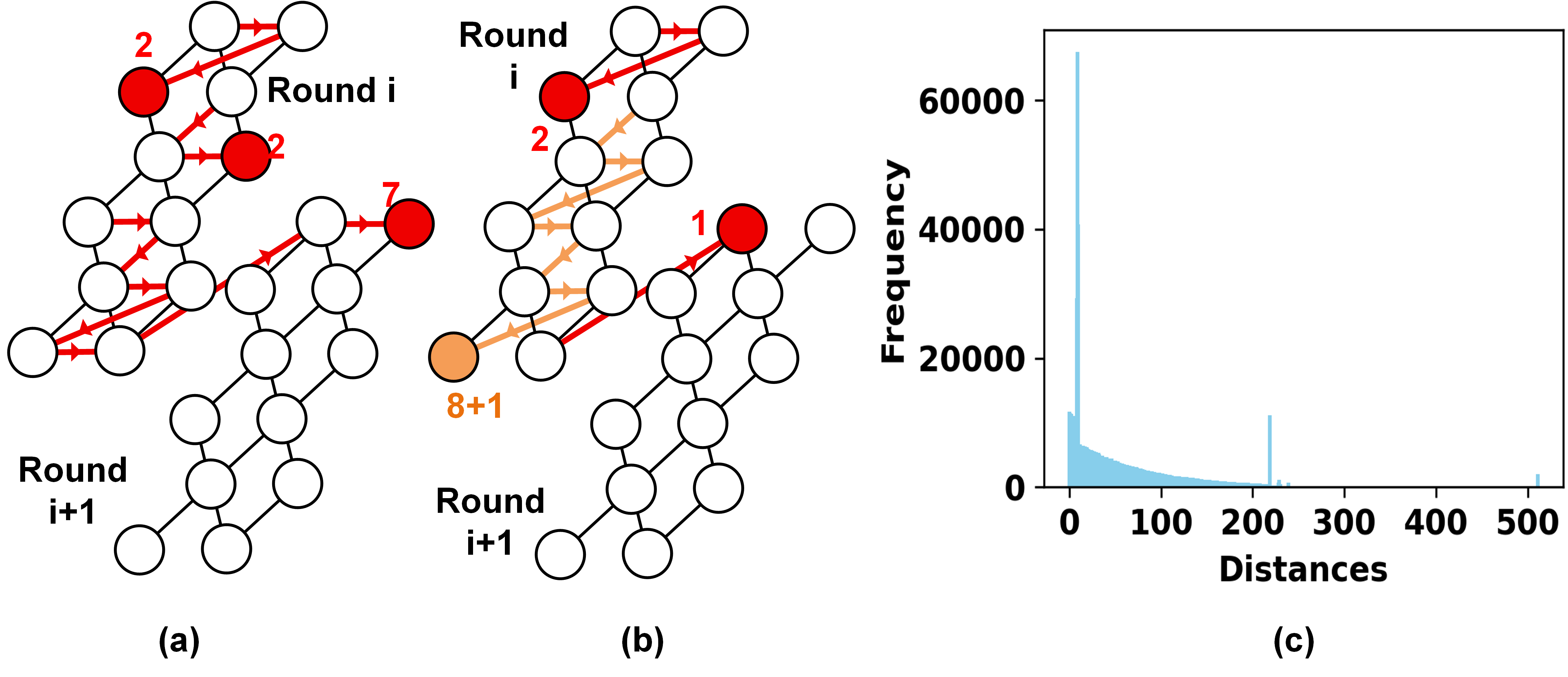}
        \vspace{-2em}
    \caption{\textit{SD} algorithm operation: (a) normal processing; (b) saturation when $max\_distance$ is reached; (c) distribution of emitted distances.}
    \label{fig:sparse}
    \vspace{-0.5cm}
\end{figure}

Leveraging the observation that syndromes remain sparse even in complex error patterns, the proposed lightweight \textit{SD} algorithm compresses the binary syndrome vector by encoding the distances between consecutive active syndromes. As illustrated in Fig. \ref{fig:sparse}(a) and (b), the first active syndrome is represented by the number of preceding zeros, while each subsequent active syndrome is represented by the number of zeros separating it from the previous one. Distances are counted continuously across syndrome round boundaries. Fig. \ref{fig:sparse}(b) highlights a special case where the maximum representable distance is reached. If the following bit is zero, a symbol corresponding to $max\_distance + 1$ is emitted. Consequently, the symbol space spans from $0$ to $max\_distance + 1$, where $0$ denotes consecutive active syndromes, $max\_distance$ corresponds to $max\_distance$ zeros followed by an active syndrome, and $max\_distance + 1$ represents the case where the maximum distance is followed by another zero.

Fig.~\ref{fig:sparse}(c) presents the distribution of distance symbols evaluated from $10^5$ blocks of $d$-round syndromes across various physical error rates and code distances, with $max\_distance$ set to 510 (resulting in valid symbol values from 0 to 511). The majority of distances are concentrated in the low range ($<256$), with occasional outliers at $512$. This trend is consistent with prior cryogenic predecoding analysis \cite{knapen2026pinball}, where the longest error chains exhibit a length of one in at least 20\% of the cases—often much more frequently. Such short error chains imply that correlated active syndromes are typically spatially close or confined within consecutive syndrome rounds, whereas uncorrelated syndromes can appear far apart, accounting for the rare 512-distance occurrences. This distribution pattern is particularly well-suited for \textit{HEnc}, which efficiently compresses frequent short-distance symbols while remaining capable of handling sparse long-distance outliers.

\subsection{Syndrome Compression Workflow}
\begin{figure}
    \centering
    \includegraphics[width=0.85\columnwidth,trim={0cm 0.3cm 0cm 0cm},clip]{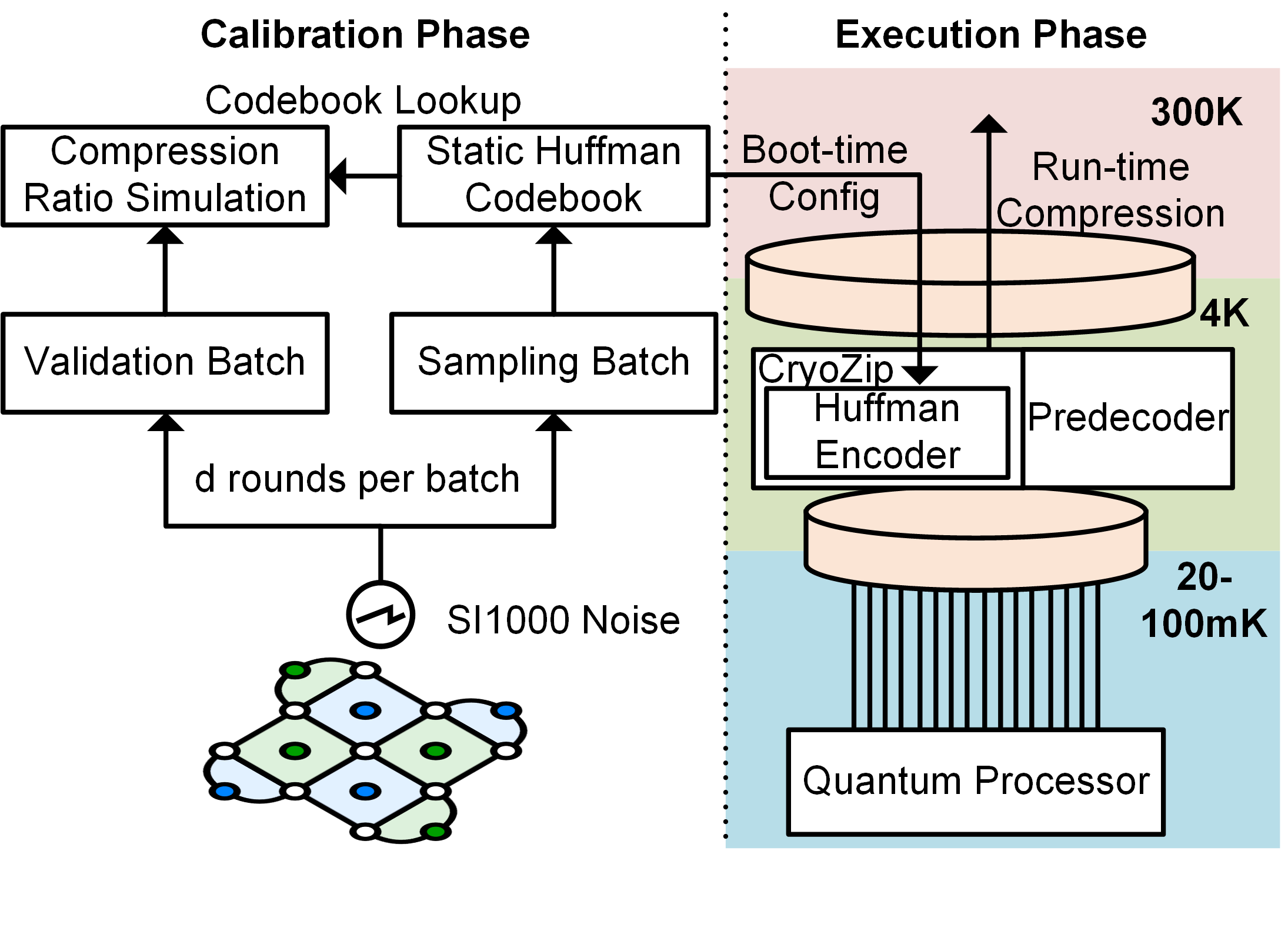}
    \caption{Workflow of CryoZip: simulation-based evaluation framework (left) and the corresponding hardware deployment flow (right).}
    \label{fig:huff}
    \vspace{-0.25cm}
\end{figure}

Next, we introduce the methodology used for the compression ratio simulation in this work and proposed hardware deployment scheme. As illustrated in Fig.~\ref{fig:huff}, during the calibration phase, $d$-round syndromes are generated using the SI1000 noise model and divided into sampling and validation batches, each containing $10^5$ blocks of $d$-round syndromes. The sampling batch is used to construct a static Huffman codebook, while the validation batch is employed to evaluate compression performance. Although the codebook remains static during the evaluation of the validation batch, it continues to perform effectively, as further discussed in Sec.~\ref{sec:evaluation}. During the hardware execution phase, this compact Huffman codebook can be scanned into the 4\,K chip during boot-time configuration and remains fixed during run-time, enabling lightweight and low-overhead binary encoding.

\section{Energy Efficient CryoZip Design}
\label{sec:hw}
We implemented compression hardware for CryoZip and three reference algorithms (DZC, GEO, and SPARSE) from AFS~\cite{das2022afs}, targeting a strict 1\,$\mu$s syndrome measurement interval~\cite{skoric2023parallel}. The hardware accepts one round of syndrome measurements at a time and outputs a valid compressed result after all $d$ rounds are processed. Because pre-processing adds latency relative to single-level decoding, the total closed-loop correction latency must remain within the reaction time introduced in Sec.~\ref{sec:transmission_overhead} and avoid backlog buildup from continuous measurements~\cite{terhal2015quantum}. To satisfy this, we conservatively budget 100\,ns per round for compression, or 10\% of the 1\,$\mu$s decoding window.

\subsection{Sliding Window Compression}
\emph{Contrary to the conclusions in}~\cite{das2022afs}---which argues that syndrome compression has negligible impact on decoding latency---we found several hardware implementation challenges in each algorithm. For DZC and GEO, $d$ rounds of syndrome measurements must be grouped before compression. Although the grouping pattern is fixed, each input bit can map to different locations across rounds, requiring substantial fan-out. Likewise, SPARSE must gather and reindex active syndromes into a continuous sequence of valid indices, creating significant fan-out and logic overhead, especially for $d>11$.

\begin{figure}
    \centering
    \includegraphics[width=\columnwidth]{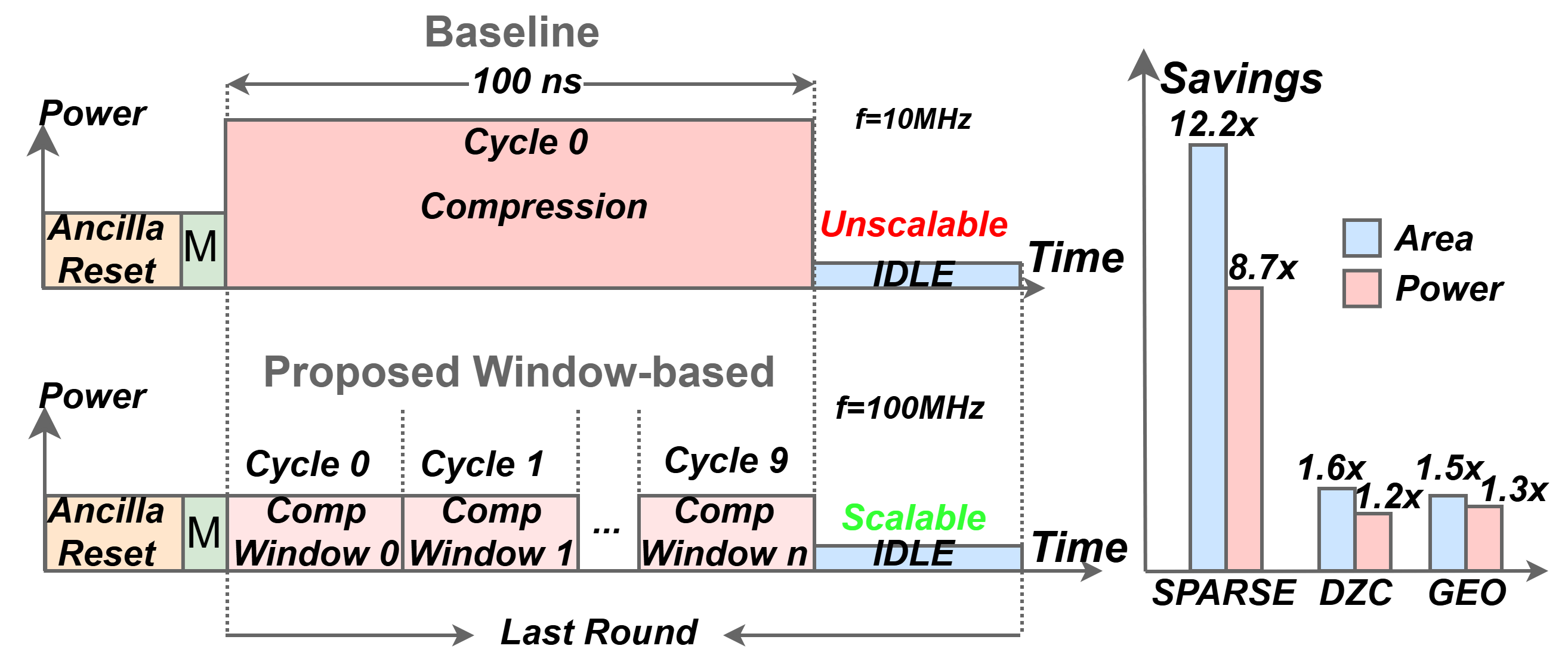}
    \caption{Sliding window compression hardware is able to achieve substantial power and area savings.}
    \label{fig:window}
    \vspace{-0.5cm}
\end{figure}

To address this, compression hardware should not process the full syndrome vector at once. Instead, our design uses a small sliding window that moves across the vector cycle by cycle, balancing area and power against modest processing latency, as shown in Fig.~\ref{fig:window}. We synthesized AFS designs with and without this technique. Sliding window compression provides substantial benefits, reducing area by up to \textbf{12.2$\times$} and power by up to \textbf{8.7$\times$}.

\begin{figure}
    \centering
    \includegraphics[width=0.85\columnwidth]{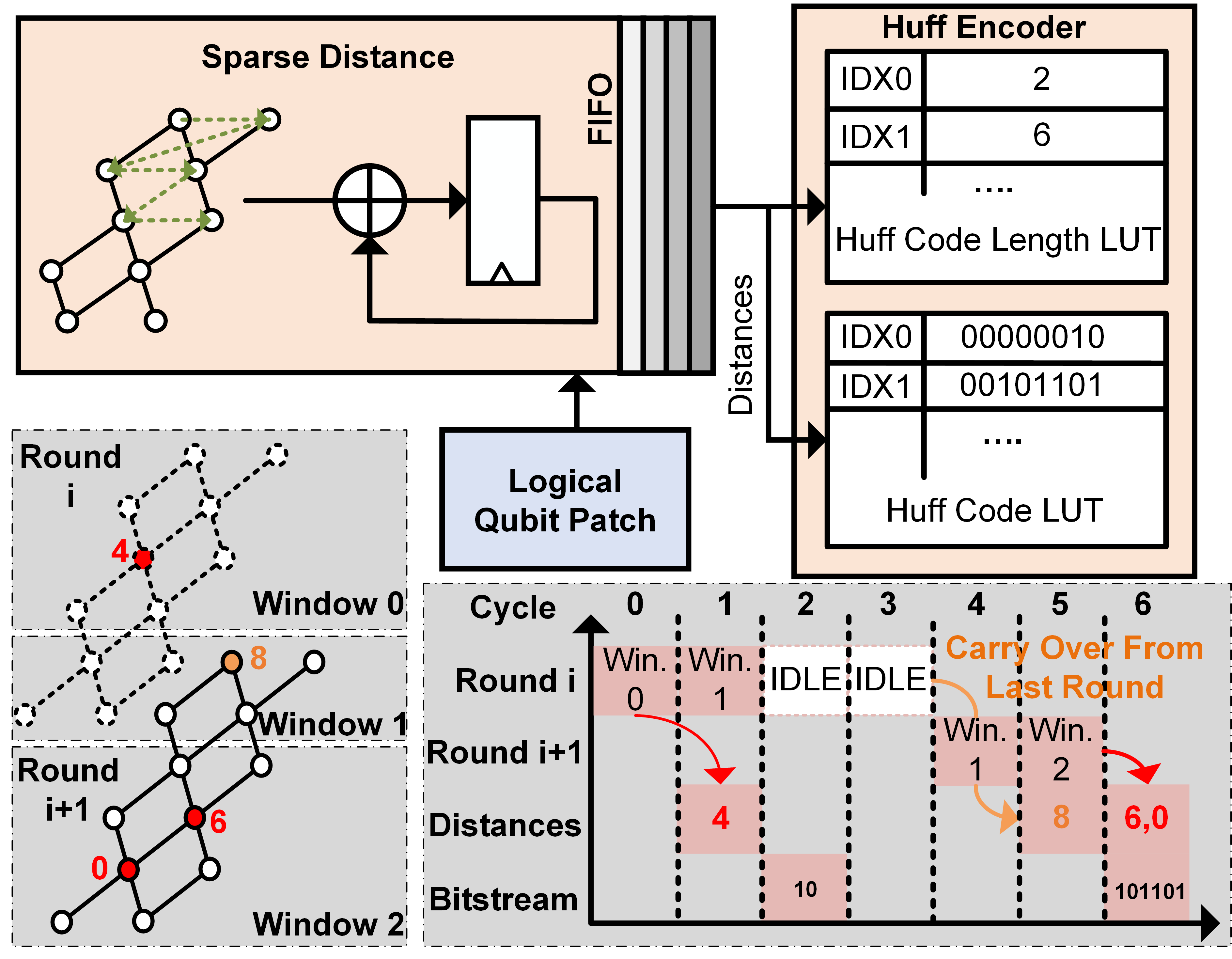}
        \vspace{-0.5em}
    \caption{Pipelined architecture of CryoZip (above) and scheduling of window-based compression (below)}
    \label{fig:arch}
    \vspace{-0.25cm}
\end{figure}

Fig.~\ref{fig:arch} (above) provides an overview of the CryoZip hardware, which comprises two main pipeline stages: \textit{SD} and \textit{HEnc}. CryoZip processes incoming syndrome measurements every 1\,$\mu$s, handling one data window per cycle. The \textit{SD} stage generates a variable number of valid distance symbols, which are buffered by a FIFO to match the input throughput of the \textit{HEnc} stage. The \textit{HEnc} stage includes two lookup tables: a \textit{Huff Code LUT} and a \textit{Huff Code Length LUT}, used to retrieve Huffman codes and their corresponding lengths. The output length from the \textit{Huff Code Length LUT} is used to clip the binary output from the \textit{Huff Code LUT}, forming a valid compressed bitstream. A downstream bit packer then concatenates these compressed binaries across $d$ measurement rounds, generating a valid output only at the final round. The resulting compressed syndrome stream is subsequently forwarded to the second-level decoder whenever the predecoder detects a complex, uncorrectable syndrome pattern.

As shown in Fig.~\ref{fig:arch} (below), although syndrome data arrives every 1\,$\mu$s, the scheme compress input data over all $d$ measurement rounds. When the $(i+1)$-th syndrome round arrives, it processes a windowed section along with any leftover data from the $i$-th round (indicated by yellow lines) to produce a valid distance output. This output is then buffered in the FIFO and later concatenated with data from previous rounds to form the final compressed stream.

\subsection{CryoZip Design Space Exploration}
\label{sec:optimization}



\begin{figure}
    \centering
    \begin{subfigure}[b]{0.49\columnwidth}
        \centering
        \includegraphics[width=\textwidth,trim={0.3cm 0.3cm 0.3cm 0.3cm},clip]{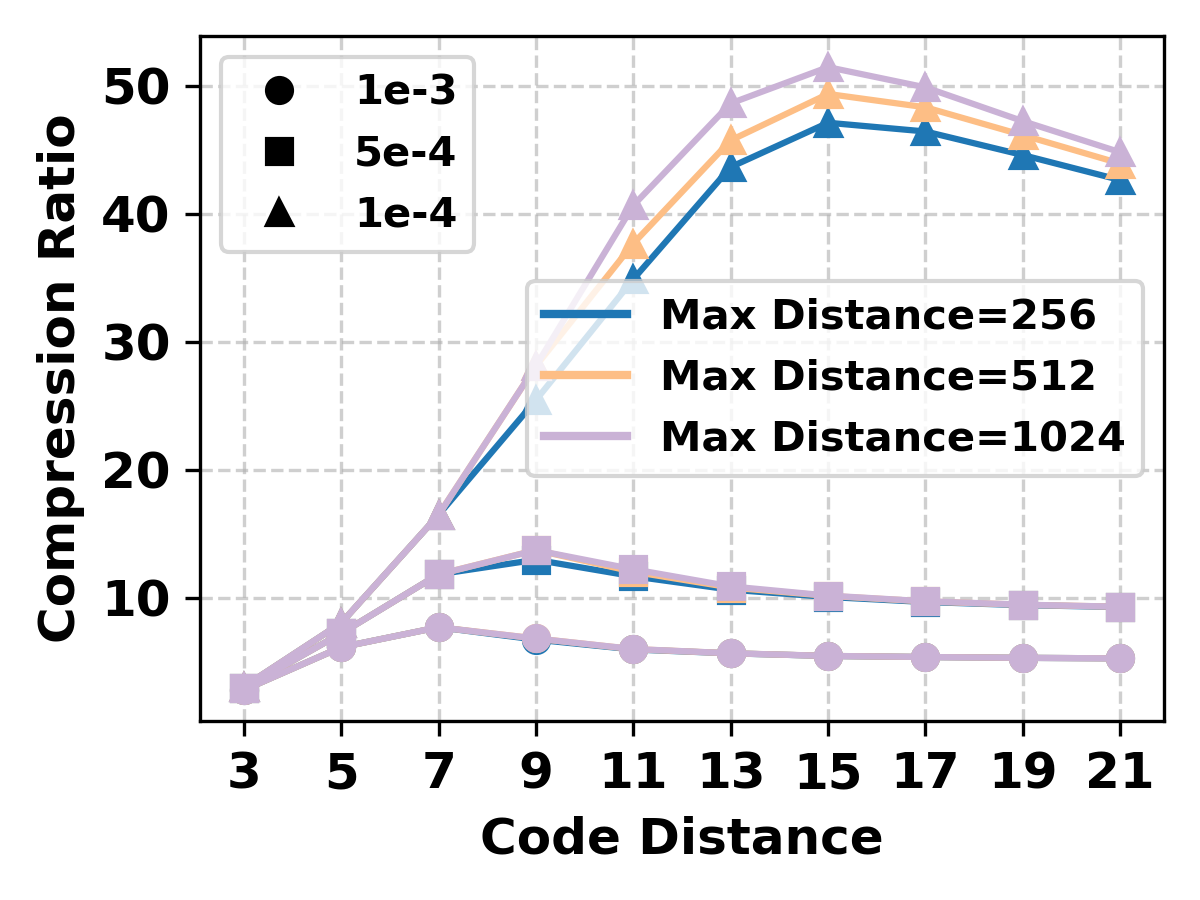}
        \caption{}
        \label{fig:tradeoff1}
    \end{subfigure}
    \begin{subfigure}[b]{0.49\columnwidth}
        \centering
        \includegraphics[width=\textwidth,trim={0.3cm 0.3cm 0.3cm 0.3cm},clip]{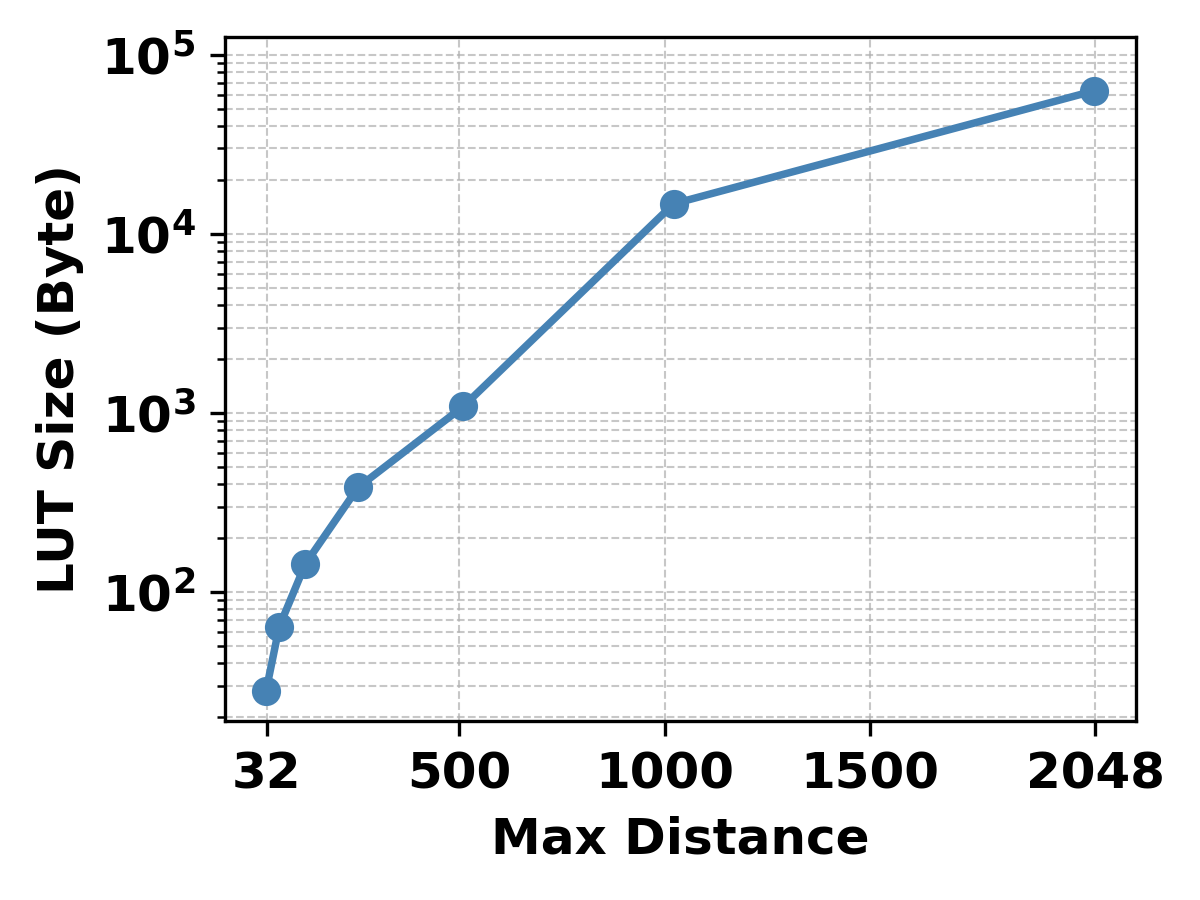}
        \caption{}
        \label{fig:tradeoff2}
    \end{subfigure}
    \vspace{-1em}
    \caption{(a) Dependence of CryoZip compression ratio on $max\_distance$. (b) LUT overhead in relative to $max\_distance$}
    \label{fig:tradeoff}
    \vspace{-0.5cm}
\end{figure}

In the \textit{SD} stage, the parameter $max\_distance$ determines both compression efficiency and \textit{HEnc} LUT size. A larger $max\_distance$ allows more consecutive zeros to be merged into single symbols, improving compression at larger code distances and lower error rates, but also increases LUT width and depth, leading to higher hardware cost. As shown in Fig.~\ref{fig:tradeoff}, the compression ratio saturates for small $d$ or high error rates, while the LUT size grows rapidly---by $13.5\times$ when $max\_distance$ increases from 512 to 1024. To balance compression gain and area overhead, we choose $max\_distance=512$ for our design.

\begin{figure}
    \centering
    \includegraphics[width=\columnwidth]{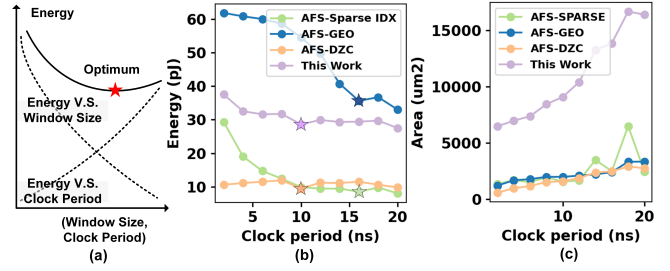}
    \vspace{-2em}
    \caption{(a) As the clock frequency decreases, the fixed-latency energy reduces, while larger window sizes increase both energy and area. The combined trend yields an optimal operating point for each design. (b) and (c) show the corresponding hardware-evaluated energy and area trends across different clock periods and window-size pairs.}
    \label{fig:opt}
    \vspace{-0.5cm}
\end{figure}

We implemented GEO, DZC, SPARSE, and CryoZip, and compared their post-synthesis power and area, as shown in Fig.~\ref{fig:opt}. Power values were obtained from vector-based simulations at 4\,K with error rate $p=1e-3$. Under a 100\,ns latency constraint (Sec.~\ref{sec:hw}) and code distance $d=21$, each design was synthesized across multiple clock frequencies by adjusting the minimum window size to meet timing.

As illustrated in Fig.~\ref{fig:opt}, lowering the clock frequency reduces fixed-latency energy, while larger window sizes increase both energy and area, resulting in an optimal operating point for each design. For CryoZip, energy decreases with longer clock periods up to 10\,ns before saturating, while area continues to grow; thus, we select 10\,ns as the optimal point (marked by stars). Similarly, the optimal points are 16\,ns for GEO, 10\,ns for DZC, and 16\,ns for SPARSE.

Although CryoZip consumes more energy than SPARSE or DZC individually, AFS runs all three algorithms in parallel with an arbiter selecting the best result. With that considered, CryoZip offers superior energy efficiency compared to AFS (Sec.~\ref{sec:esave}). Its slightly larger area, mainly from SRAM-based LUTs in the \textit{HEnc} stage, can be amortized across multiple logical qubits. Moreover, in cryogenic environments, area is a secondary concern compared to power and bandwidth constraints.

\section{Evaluation}
\label{sec:evaluation}

\begin{figure*}[!htb]
    \centering
    \begin{subfigure}{0.49\textwidth}
        \centering
        \includegraphics[width=\textwidth,trim={0.3cm 0.3cm 0.3cm 0.3cm},clip]{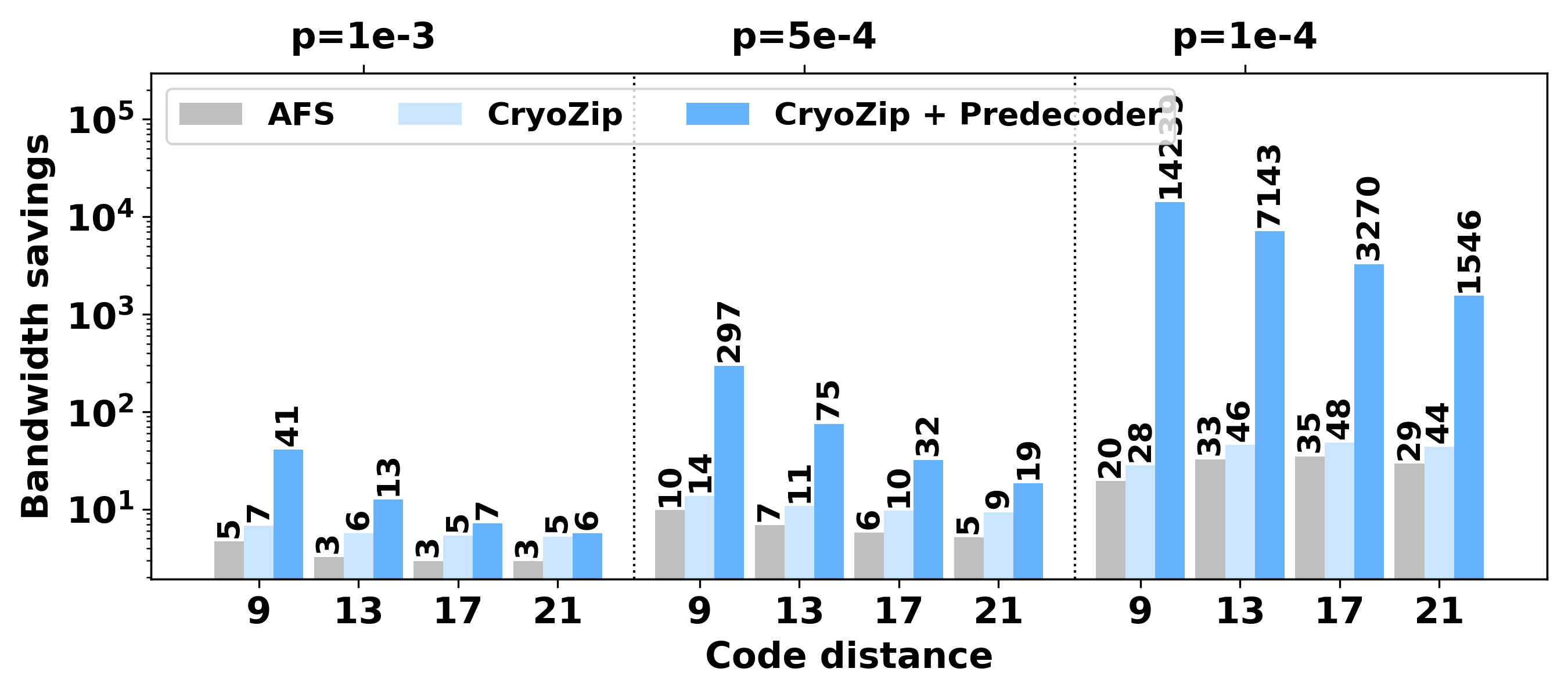}
        \caption{}
        \label{fig:bsave}
    \end{subfigure}
    \begin{subfigure}{0.49\textwidth}
        \centering
        \includegraphics[width=\textwidth,trim={0.3cm 0.3cm 0.3cm 0.3cm},clip]{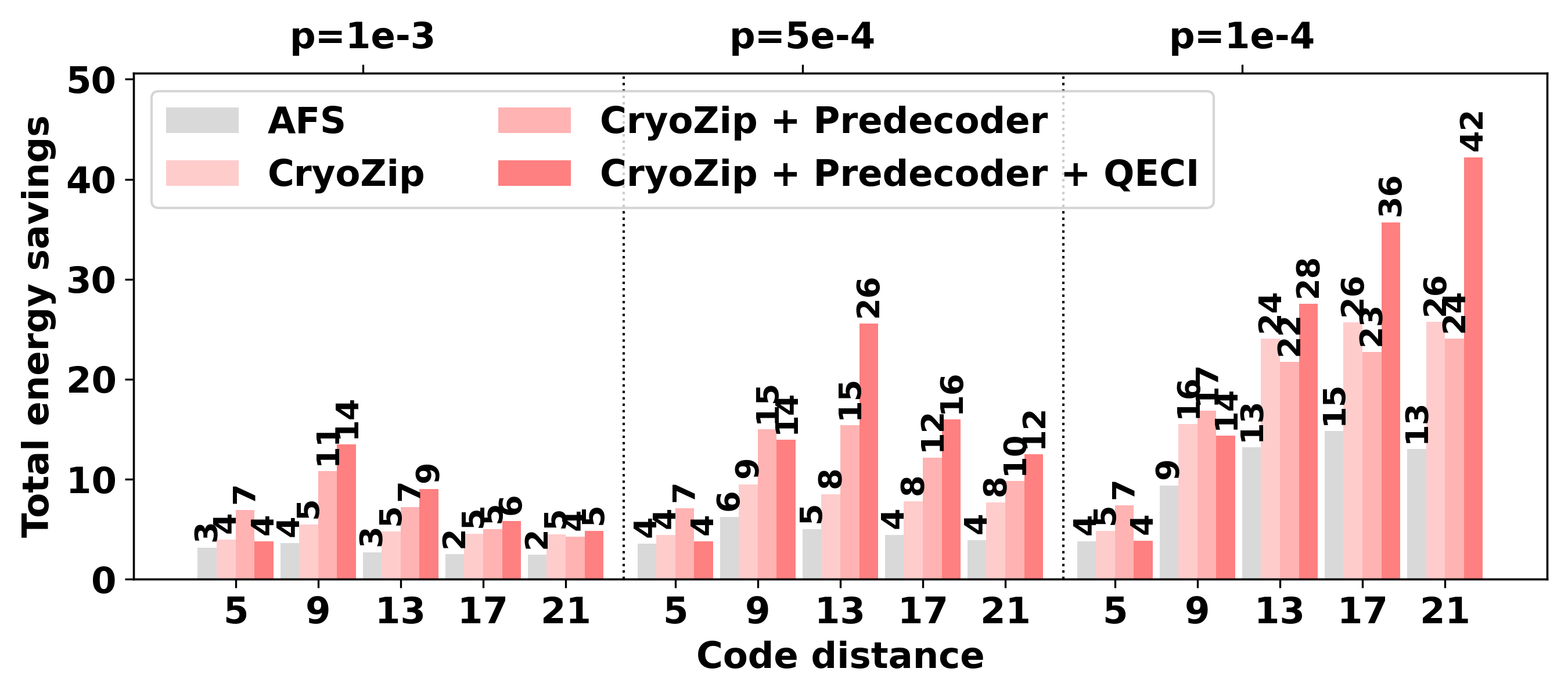}
        \caption{}
        \label{fig:esave}
    \end{subfigure}
    \vspace{-1.25em}
    \caption{(a) Bandwidth savings on the surface code for AFS, CryoZip, and CryoZip with the predecoder; (b) Total energy savings combines the benefits of CryoZip and predecoding.}
    \label{fig:ebsave}
    \vspace{-0.25cm}
\end{figure*}

We evaluate the AFS \cite{das2022afs} and CryoZip compression algorithms across multiple QEC codes—--surface, BB, and color—--that balance compression effectiveness on sparse data with hardware efficiency. As described in Sec.~\ref{sec:optimization}, each design is evaluated at its most energy-efficient operating point. Under this fair comparison, CryoZip consistently outperforms AFS, achieving comparable or lower energy consumption while delivering significantly higher compression ratios (Sec.~\ref{sec:cr}) and superior overall energy savings (Sec.~\ref{sec:esave}).

\subsection{Compression Ratio}
\label{sec:cr}
\begin{figure}
    \centering
    \begin{subfigure}{0.32\columnwidth}
        \centering
        \includegraphics[width=\textwidth,trim={0.3cm 0.3cm 0.3cm 0.3cm},clip]{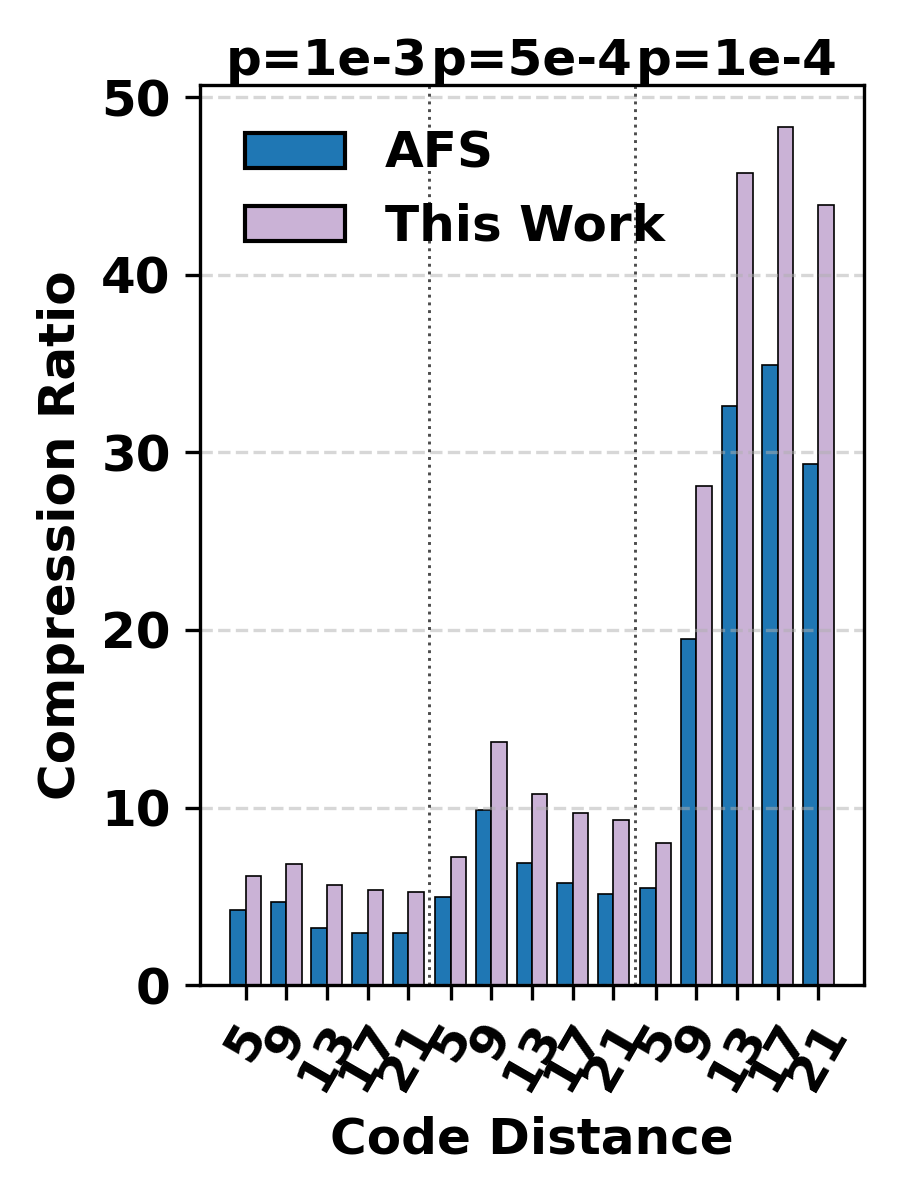}
        \caption{}
        \label{fig:cr_sc}
    \end{subfigure}
    \begin{subfigure}{0.32\columnwidth}
        \centering
        \includegraphics[width=\textwidth,trim={0.3cm 0.3cm 0.3cm 0.3cm},clip]{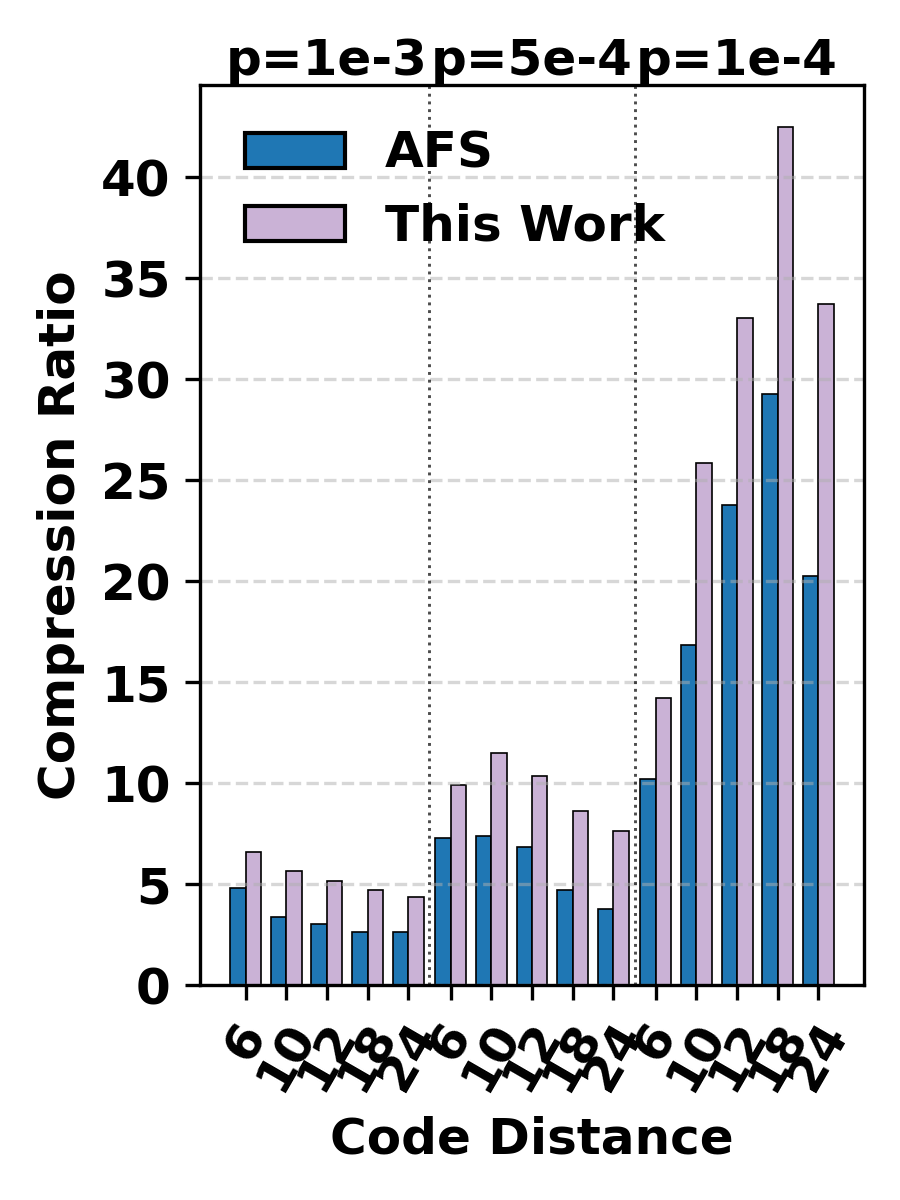}
        \caption{}
        \label{fig:cr_bb}
    \end{subfigure}
    \begin{subfigure}{0.32\columnwidth}
        \centering
        \includegraphics[width=\textwidth,trim={0.3cm 0.3cm 0.3cm 0.3cm},clip]{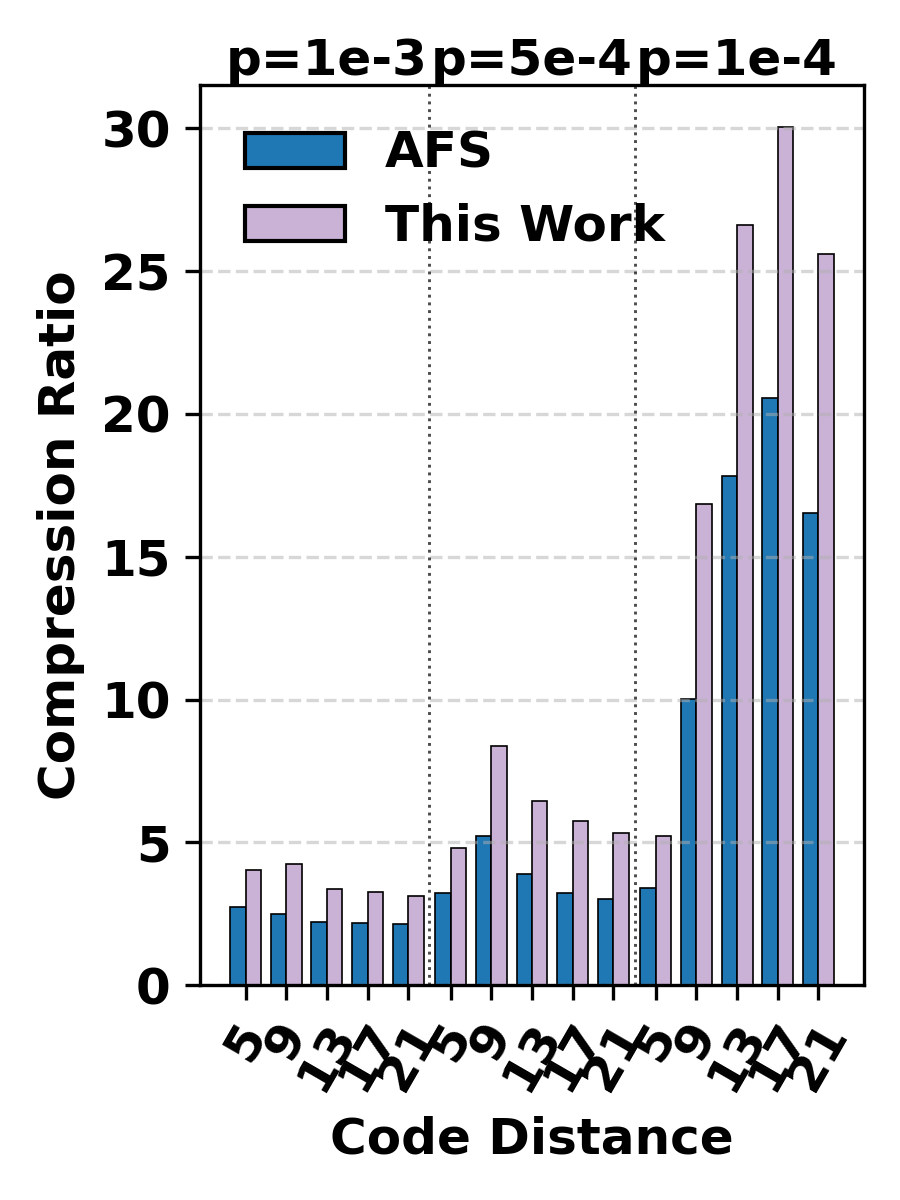}
        \caption{}
        \label{fig:cr_cc}
    \end{subfigure}
    \vspace{-1.25em}

    \caption{Compression ratio comparison of AFS and CryoZip on (a) surface code; (b) BB code; (c) color code.}
    \label{fig:cr}
    \vspace{-0.5cm}
\end{figure}

Fig.~\ref{fig:cr} presents the compression ratios of each algorithm on the three codes. For surface and color codes with distance $d$, we compress $d$ rounds of $O(d^2)$ syndrome data from one logical qubit per decoding cycle. For the BB code, we use configurations $d=6,10,12,18,24$ from \cite{bravyi2024high}, which encode multiple logical qubits. Since GEO is tailored to surface codes, it is excluded for other codes. Each algorithm is tested on 10,000 instances (excluding trivial cases that can be resolved by predecoder), and the reported \textit{compression ratio} is the average per $d$ rounds.


We show how the compression ratio varies with code distance and physical error rates ($1e-3$, $5e-4$, $1e-4$) across the three QEC codes. The baseline AFS ratio is determined as the best among GEO, DZC, and SPARSE configurations. Larger $d$ and lower error rates produce sparser syndromes, resulting in higher compression ratios. Overall, qLDPC codes such as BB (1.91--42.5$\times$) and color (1.41--30.0$\times$) exhibit lower ratios than the surface code (1.50--48.3$\times$) due to their denser structure. CryoZip consistently surpasses all algorithms from AFS, delivering up to 1.81$\times$ higher compression. For the surface code [Fig.~\ref{fig:cr}(a)], CryoZip achieves 48.3$\times$ at $d=17$, $p=0.1e-4$, and 5.35$\times$ at $d=21$, $p=1e-3$. For the BB code [Fig.~\ref{fig:cr}(b)], the peak ratio is 42.46$\times$ and the lowest 4.38$\times$. For the color code [Fig.~\ref{fig:cr}(c)], CryoZip reaches 30.03$\times$, with a minimum of 3.21$\times$.

GEO and DZC perform well under higher error rates and larger code distances, since many new non-zero syndromes fall within existing non-zero groups, keeping the compressed size relatively stable. However, their efficiency is limited by the fact that even a single non-zero element within a group prevents that group from being compressed effectively. SPARSE, on the other hand, encodes every non-zero bit individually using fixed-length binary indices, which restricts its compression capability. In contrast, CryoZip encodes the distance between consecutive non-zero bits rather than their absolute indices. This approach offers two key benefits: distances are smaller in magnitude and their distribution is more compact. CryoZip further exploits this compact distribution through entropy coding, assigning shorter codes to more frequent distances instead of fixed bitwidth representations, achieving the highest overall compression efficiency among all algorithms.

\vspace{-0.5em}
\subsection{Bandwidth Savings}


We illustrate the syndrome bandwidth savings on the surface code achieved by CryoZip, and by CryoZip in cooperation with the predecoder, in Fig.~\ref{fig:ebsave} (a). CryoZip alone delivers up to 48$\times$ bandwidth reduction and maintains at least 5$\times$ reduction even in the worst case. Regardless of the underlying physical error rate $p$, CryoZip with the predecoder provides substantial bandwidth reduction: for $d \leq 5$, predecoder achieves nearly full coverage, eliminating most transmissions. At $d=9$ and $p=1e-4$, combining CryoZip with the predecoder yields a maximum bandwidth reduction of 14,239$\times$. Even in the worst case ($d=21$, $p=1e-3$), a $6\times$ reduction is still achieved. 

\vspace{-0.5em}
\subsection{Energy Savings}
\label{sec:esave}


We evaluate the energy of CryoZip, CryoZip combined with the predecoder, and the AFS system, accounting for both hardware power and 4\,K-to-RT transmission energy \cite{fakkel2024cryo}. As shown in Fig.~\ref{fig:ebsave} (b), while the hardware overhead of the predecoder and CryoZip increases with code distance and error rate, their combined efficiency consistently outperforms the baseline due to CryoZip’s high compression ratio and the predecoder’s high coverage rate.

At $p=1e-3$ (eFTQC regime), CryoZip achieves $4$-$5\times$ energy savings, and CryoZip with predecoder achieves $4$-$11\times$. For practical eFTQC systems ($\sim$$10^4$ physical qubits, $d<11$), savings are $4$-$5\times$ and $7$-$11\times$, respectively. As error rates decrease, energy savings increase: $7$-$15\times$ at $p=5e-4$ and $7$-$24\times$ at $p=1e-4$. Thus, CryoZip remains scalable and energy-efficient for higher code distances in future FTQC systems.

These results conservatively assume only syndrome payloads are transmitted. In realistic protocols like Riverlane’s QECi \cite{riverlane2024qeci}, 64-bit packets with 32-bit headers double transmission costs, further boosting CryoZip with predecoder’s energy savings up to $42\times$ at $p=1e-4$.

\vspace{-0.5em}
\section{Conclusion}

We introduced CryoZip, a cryogenic compression framework that cooperates with lightweight predecoders to reduce syndrome transmission under realistic, circuit-level noise across multiple QEC codes. Our sliding-window architecture translates QEC latency constraints into an energy-efficient design that fits within cryogenic power budgets. A 22\,nm FDSOI implementation at 4\,K demonstrate that algorithm–architecture–technology co-design can make syndrome handling practical at scale. Together, these results form a path toward bandwidth and energy-efficient QEC pipelines. CryoZip is the first scalable cryogenic compressor that sustains high compression ratios across multiple QEC codes under noise representative of real quantum systems.

%
\begin{acks}
This material is based upon work supported by the U.S. Department of Energy, Office of Science, Office of Advanced Scientific Computing Research, Accelerated Research in Quantum Computing under Award Number DE-SC0025633. This research used resources of the National Energy Research Scientific Computing Center, a DOE Office of Science User Facility supported by the Office of Science of the U.S. Department of Energy under Contract No. DE-AC02-05CH11231 using NERSC award NERSC DDR-ERCAP0035341.
This research is supported in part by funding from the Quantum Research Institute at the University of Michigan.
The authors also thank Semiwise Ltd., UK for access to the cryo-CMOS PDK used for hardware evaluation in this work.
\end{acks}

\bibliographystyle{ACM-Reference-Format}
\bibliography{refs}


\end{document}